\documentclass[12pt]{article}  
\usepackage[dvips]{graphicx}

\def\beq{\begin{equation}}
\def\eeq{\end{equation}}
\def\to{\rightarrow}

\def\bsg{\ifmmode B\to X_s\gamma\else $B\to X_s\gamma$\fi}
\def\bsll{\ifmmode B\to X_s\ell^+\ell^-\else $B\to X_s\ell^+\ell^-$\fi}
\def\bstt{\ifmmode B\to X_s\tau^+\tau^-\else $B\to X_s\tau^+\tau^-$\fi}
\def\shat{\ifmmode \hat{s}\else $\hat{s}$\fi}

\newcommand{\newc}{\newcommand}

\newc{\lcal}{\int {\cal L}dt}
 
\newc{\LSP}{{\chi^0_1}}
\newc{\stauR}{{\tilde \tau_R}}
\newc{\stau}{{\tilde \tau_1}}
\newc{\mstop}{m_{\tilde{t}}}
\newc{\mHpm}{m_{H^\pm}}
\newc{\gsim}{\lower.7ex\hbox{$\;\stackrel{\textstyle>}{\sim}\;$}}
\newc{\lsim}{\lower.7ex\hbox{$\;\stackrel{\textstyle<}{\sim}\;$}}
\newc{\ie}{{\it i.e.}}          
\newc{\etal}{{\it et al.}}
\newc{\eg}{{\it e.g.}}          
\newc{\kev}{\hbox{\rm\,keV}}            
\newc{\mev}{\hbox{\rm\,MeV}}            
\newc{\gev}{\hbox{\rm\,GeV}}            
\newc{\tev}{\hbox{\rm\,TeV}}
\newc{\xpb}{\hbox{\rm\, pb}}
\newc{\xfb}{\hbox{\rm\, fb}}

\newc{\mtop}{m_t}
\newc{\mbot}{m_b}
\newc{\mz}{M_Z}
\newc{\mw}{M_W}
\newc{\alphasmz}{\alpha_s(M_Z)}
\newc{\swsq}{\sin^2\theta_W}
\newc{\cwsq}{\cos^2\theta_W}
\newc{\tw}{\tan\theta_W}
\newc{\cw}{\cos\theta_W}
\newc{\sw}{\sin\theta_W}
\newc{\BR}{\hbox{\rm BR}}
\newc{\zbb}{Z\to b\bar}
\newc{\Gb}{\Gamma (Z\to b\bar b)}
\newc{\Gh}{\Gamma (Z\to \hbox{\rm hadrons})}
\newc{\rbsm}{R_b^\hbox{\rm sm}}
\newc{\rbsusy}{R_b^\hbox{\rm susy}}
\newc{\drb}{\delta R_b}

\newc{\sgn}{\mbox{sgn}}

\def\eq#1{eq.~(\ref{#1})}
\def\fig#1{fig.~\ref{#1}}
\def\beqa{\begin{eqnarray}}
\def\eeqa{\end{eqnarray}}

\def\gtilu{{\tilde g}_u}
\def\gtild{{\tilde g}_d}
\def\gtilup{{\tilde g}_u^\prime}
\def\gtildp{{\tilde g}_d^\prime}
\def\gtiluq{{\tilde g}_u^2}
\def\gtildq{{\tilde g}_d^2}
\def\gtilupq{{\tilde g}_u^{\prime 2}}
\def\gtildpq{{\tilde g}_d^{\prime 2}}

\def\sps{Split Supersymmetry}
\def\mtil{\tilde{m}}

\def\beq{\begin{equation}}
\def\eeq{\end{equation}}
\def\bea{\begin{eqnarray}}
\def\eea{\end{eqnarray}}
\def\slashchar#1{\setbox0=\hbox{$#1$}           
   \dimen0=\wd0                                 
   \setbox1=\hbox{/} \dimen1=\wd1               
   \ifdim\dimen0>\dimen1                        
      \rlap{\hbox to \dimen0{\hfil/\hfil}}      
      #1                                        
   \else                                        
      \rlap{\hbox to \dimen1{\hfil$#1$\hfil}}   
      /                                         
   \fi}                                         %
\catcode`@=11
\long\def\@caption#1[#2]#3{\par\addcontentsline{\csname
  ext@#1\endcsname}{#1}{\protect\numberline{\csname
  the#1\endcsname}{\ignorespaces #2}}\begingroup
    \small
    \@parboxrestore
    \@makecaption{\csname fnum@#1\endcsname}{\ignorespaces #3}\par
  \endgroup}
\catcode`@=12

\begin{document}

\baselineskip=18pt

\setcounter{footnote}{0}
\setcounter{figure}{0}
\setcounter{table}{0}

\begin{titlepage}
June 2005 \hspace*{\fill}
CERN-TH/2005-108

\begin{center}
\vspace{1cm}

{\Large \bf On the Tuning Condition of Split Supersymmetry}

\vspace{0.8cm}

{\bf A. Delgado, G.F. Giudice} 

\vspace{.5cm}

{\it CERN, Theory Division, CH-1211 Geneva 23, Switzerland}

\end{center}
\vspace{1cm}

\begin{abstract}
\medskip
Split Supersymmetry does not attempt to solve the hierarchy problem,
but it assumes a tuning condition for the electroweak scale. We 
clarify the meaning of this condition and show how it is related to
the underlying parameters. Simple assumptions on the structure of
the soft terms lead to predictions on $\tan \beta$ and on the physical
Higgs mass.

\end{abstract}

\bigskip
\bigskip


\end{titlepage}



\sps~\cite{savas,split,noi4} does not attempt to give a natural explanation
of the hierarchy problem but, as in the Standard Model, a parameter 
fine-tuning is assumed to obtain the
correct breaking of electroweak symmetry. In this paper we want to
show that, although the theory does
not provide a dynamical explanation for the tuning, its existence
leads to important information on the underlying parameters and on
measurable physical quantities.

Let us start by considering the potential in \sps\ 
for the real and neutral Higgs
field $h$, valid at energies lower than the squark and slepton mass
scale $\mtil$,
\beq
V=\frac{m^2}{2}h^2+\frac{\lambda}{ 8}h^4.
\label{pot}
\eeq
The Higgs mass parameter $m^2$ satisfies to the RG equation
\bea
&&16\pi^2{\bar \mu}\frac{d}{d {\bar \mu}}
~ m^2=\left[ 6\lambda -\frac{3}{2} 
\left( g^{\prime 2} +3g^2\right) +3\left( \gtiluq +\gtildq \right)
+\gtilupq +\gtildpq +6 h_t^2 \right] m^2 \nonumber \\
&& -2\left[ \left( \gtilupq +\gtildpq \right) \left( M_1^2 +\mu^2\right)
+3 \left( \gtiluq +\gtildq \right) \left( M_2^2 +\mu^2\right)
+2 \left( \gtilup \gtildp M_1 + 3 \gtilu \gtild M_2 \right) \mu \right] .
\label{run}
\eea
Here $M_{1,2}$ and $\mu$ are the gaugino and higgsino masses, 
$h_t$ is the top-Yukawa coupling, and ${\tilde g}_{u,d}$,
${\tilde g}^\prime_{u,d}$ are the gaugino-higgsino Yukawa couplings
defined as in ref.~\cite{split}. Notice that the term proportional
to $M_{1,2}$ and $\mu$ always gives a negative contribution to 
\eq{run}. Therefore, if $m^2$ starts positive at the scale $\mtil$,
it will remain positive as the renormalization scale $\bar \mu$ is
lowered.
Thus below $\mtil$, in the energy range of the Split-Supersymmetry
effective theory, it is {\it not} possible to obtain radiative
electroweak breaking. 
Had gaugino and higgsino masses induced the radiative breaking, we could
have hoped to relate the dark-matter mass scale ($M_{1,2}$, $\mu$) to
the electroweak scale ($-m^2/\lambda$), providing a connection that is
still missing in the context of {\sps}.   

The negative result on radiative breaking from gaugino and higgsino
masses can be generalized. The RG evolution of the Higgs mass parameter
$m^2$, for a generic field content, can be obtained by requiring that
the one-loop effective potential is independent of the renormalization
scale,
\beq
16\pi^2{\bar \mu}\frac{d}{d {\bar \mu}}~ m^2 =2 \gamma_hm^2 +\Delta 
\label{genrun}
\eeq
\beq
\Delta =
\frac{1}{2}
\frac{d^2}{dh^2}\left. {\rm STr}\left({\cal M}^\dagger
{\cal M}\right)^2\right|_{h=0}.
\label{delt}
\eeq
Here $\gamma_h=-16\pi^2({\bar \mu} /h)(dh/d{\bar \mu})$ is the Higgs anomalous dimension
and ${\cal M}$ is the mass matrix of all particles in the Higgs background.
The term proportional to $\gamma_h$ in \eq{genrun} gives a {\it multiplicative}
renormalization of $m^2$ and therefore cannot reverse the sign of $m^2$
during the RG evolution, and cannot
induce radiative breaking. The second term in \eq{genrun} can give an
{\it additive} renormalization and, if positive, can trigger 
spontaneous symmetry
breaking. For chiral
fermions linearly coupled to the Higgs, we have ${\cal M}=A+Bh$ (with 
generic matrices $A$ and $B$). This gives a contribution to $\Delta$
\beq  
\Delta = -2 {\rm Tr} \left( X^\dagger X +Y^\dagger Y\right) ,
\label{matr}
\eeq
\beq
X\equiv AB^\dagger +BA^\dagger,~~~~Y\equiv \sqrt{2} A^\dagger B .
\eeq
It is apparent that \eq{matr} always gives a negative contribution,
and therefore fermions linearly-coupled to the Higgs cannot trigger
radiative breaking. 

Fermions are actually responsible for radiative breaking in Little-Higgs 
theories~\cite{little},
but this is not in contradiction with our result. In Little-Higgs models,
there are non-renormalizable couplings between the new fermions and Higgs
bilinears, of the form  ${\cal M}=A+Bh^2$, which can lead to positive values
of $\Delta$, for appropriate values of $A$ and $B$, see \eq{delt}. In
{\sps}, we can obtain couplings of the form ${\cal M}=A+Bh^2$, considering
the case $M_{1,2}\gg \mu$ (or viceversa). Integrating out the gauginos,
we find an effective 
higgsino coupling to Higgs bilinears which, through \eq{genrun},
gives $m^2 \propto (\alpha /\pi) (\mu^3/M) \ln (M/\mu)$. For an appropriate 
choice of the sign of the higgsino mass $\mu$, we obtain a negative
contribution to the Higgs mass parameter $m^2$. However, in 
realistic models, one cannot explain
why other, parametrically larger, effects (such as
$m^2 \propto (\alpha /\pi) M^2 \ln (\mtil /M)$, not to mention the boundary
condition of $m^2$ at the scale $\mtil$) are smaller in size than the
negative contribution from higgsinos.

On the other hand, 
above the scale $\mtil$, the radiative breaking is easily achieved~\cite{rad} 
by the positive contribution to $\Delta$ from the stop  
\beq
\Delta_{\tilde t} = 6 h_t^2 (\mtil_{Q_3}^2+\mtil_{U_3}^2),
\eeq
where $\mtil_{Q_3,U_3}$ are the ${\tilde t}_{L,R}$ soft masses, of the
order of $\mtil$.
This leads to $m^2 ={\cal O}(-\mtil^2)$, which has to be tamed by
the tuning condition. From our discussion, it is now manifest 
the meaning of the
tuning: {\it it corresponds to the condition that the 
Higgs mass parameter $m^2$
changes sign precisely at the scale $\mtil$, 
at which the squarks
are integrated out.} 
As shown before, the running below $\mtil$ has a 
modest impact on $m^2$ and, in particular, it does not further change its sign.

The supersymmetric Higgs potential above $\mtil$, 
along the real and neutral components, is
\beq
V=\frac{m_{H_u}^2}{2}H_u^2 + \frac{m_{H_d}^2}{2}H_d^2 -B_\mu H_u H_d
+\frac{g^2+g^{\prime 2}}{32}\left( H_u^2-H_d^2 \right)^2,
\eeq
with mass parameters of the order of $\mtil^2$.
It is convenient to rewrite the potential in terms of the fields
\beq
\pmatrix{h\cr H} = \pmatrix{\cos\beta & \sin\beta \cr -\sin\beta &\cos\beta}
\pmatrix{H_d \cr H_u} ,~~~~\sin 2\beta \equiv \frac{2B_\mu}{m_{H_u}^2
+m_{H_d}^2}.
\eeq
We obtain
\beq
V=\frac{\lambda}{8}\left( h^2+2\frac{m^2}{\lambda}-2\tan 2\beta ~hH -H^2
\right)^2 + \frac{m_{H_u}^2
+m_{H_d}^2}{2}H^2,
\label{pots}
\eeq
\beq
\lambda = \frac{g^2+g^{\prime 2}}{4}\cos^22\beta ,~~~~m^2=\frac{(t^2-1)}
{(t^2+1)^2}\left( m_{H_u}^2 t^2 -m_{H_d}^2\right) ,~~~~t\equiv \tan\beta .
\label{bou}
\eeq
For $m_{H_u}^2+m_{H_d}^2={\cal O}(\mtil^2)$, the field $H$ decouples
at the squark-mass scale. The tuning condition corresponds to choosing
$m^2$
approximately zero (and slightly negative). Once this condition is applied,
the potential of the effective theory coincides with
\eq{pot}, with the boundary conditions at $\mtil$ given by \eq{bou}.

\begin{table}
\centering
\begin{tabular}{|c|c|c|}
\hline
$\mtil$(GeV) & $K\sin^2\beta$ & $\omega$ \\
\hline
\hline
$10^4$& $0.28$&$0.024$\\
$10^7$& $0.19$&$0.020$\\
$10^{10}$& $0.12$&$0.016$\\
$10^{13}$& $0.06$&$0.009$\\
\hline
\end{tabular}
\caption{Values of $K\sin^2\beta$ (this combination is almost independent of 
$\tan\beta$) and $\omega$ for different values of $\mtil$ and for
$m_t^{\rm pole}=178$~GeV.}
\label{tab}
\end{table}

We want to investigate what information the tuning condition can provide us
on the fundamental parameters at a large scale $M_X$, which we will
identify with the GUT scale. The relations between the soft masses
at the scale $M_X$ (denoted by a bar) and those at the matching scale
$\mtil$ are
\bea
m_{H_u}^2(\mtil ) &=& {\overline m}_{H_u}^2-K\left( {\overline m}_{H_u}^2
+{\overline m}_{Q_3}^2+{\overline m}_{U_3}^2\right) -\omega {\overline S}\\
m_{H_d}^2(\mtil ) &=& {\overline m}_{H_d}^2
+\omega {\overline S}\\
m_{Q_3}^2(\mtil ) &=& {\overline m}_{Q_3}^2-\frac{K}{3}
\left( {\overline m}_{H_u}^2
+{\overline m}_{Q_3}^2+{\overline m}_{U_3}^2\right) -\frac{\omega}{3} 
{\overline S}\\
m_{U_3}^2(\mtil ) &=& {\overline m}_{U_3}^2-\frac{2K}{3}
\left( {\overline m}_{H_u}^2
+{\overline m}_{Q_3}^2+{\overline m}_{U_3}^2\right) +\frac{4\omega}{3} 
{\overline S}\\
{\overline S}&\equiv& {\overline m}_{H_u}^2-{\overline m}_{H_d}^2
+\sum_{i=1}^3 \left( {\overline m}_{Q_i}^2 -2{\overline m}_{U_i}^2
+{\overline m}_{D_i}^2-{\overline m}_{L_i}^2+{\overline m}_{E_i}^2\right) .
\eea
Here we have neglected terms of weak-scale size ($M$, $\mu$, $A$) and
kept only the one-loop leading contributions from gauge and top-Yukawa
couplings (which is adequate for not too large values of $\tan\beta$). 
The coefficients $K$ and $\omega$ are given by
\beq
K=\left. \frac{3F\lambda_t^2}{16\pi^2E}\right|_{\mtil}~~~~
E({\bar \mu})=\prod_{i=1}^3\left[ \frac{g_i^2(M_X)}{g_i^2({\bar \mu})}
\right]^{\frac{2c_i}
{b_i}} ~~~~ F=\int_0^{\ln M_X^2/\mtil^2}d(\ln M_X^2/{\bar \mu}^2)~E
\label{cappa}
\eeq
\beq
\omega = \frac{3g_1^2(\mtil)}{160\pi^2}\ln \frac{M_X^2}{\mtil^2},
\eeq
where $c_i=(13/30,3/2,8/3)$, $b_i=(33/5,1,-3)$, and $g_1$ has GUT 
normalization. The numerical values of $K$ and $\omega$ for different
$\mtil$ are given in table~1. In \eq{cappa} 
$\lambda_t$ is the supersymmetric top-Yukawa coupling
related to the low-energy coupling $h_t$ by the matching condition
$\lambda_t(\mtil)\sin\beta=h_t(\mtil)$. The parameter $K$ has a maximum
value ($K<1/2$), once we require the absence of Landau poles below the scale 
$M_X$.

The conditions for tuning the electroweak scale are
\bea
&&m^2(\mtil)\equiv \frac{(t^2-1)}
{(t^2+1)^2}\left[ m_{H_u}^2(\mtil) t^2 -m_{H_d}^2(\mtil)\right] =0
\label{tun1}\\
&&m_{H_u}^2(\mtil)+m_{H_d}^2(\mtil)>0\label{tun2}\\
&&m_{Q_3}^2(\mtil)>0\label{tun3}\\
&&m_{U_3}^2(\mtil)>0 .\label{tun4}
\eea
Equations~(\ref{tun2})--(\ref{tun4}) correspond to the requirement that
no electric-charge or colour breaking minima are developed at the scale
$\mtil$, as recently discussed in ref.~\cite{ibarra}.

\begin{figure}
\centering
\includegraphics[width=0.48\linewidth]{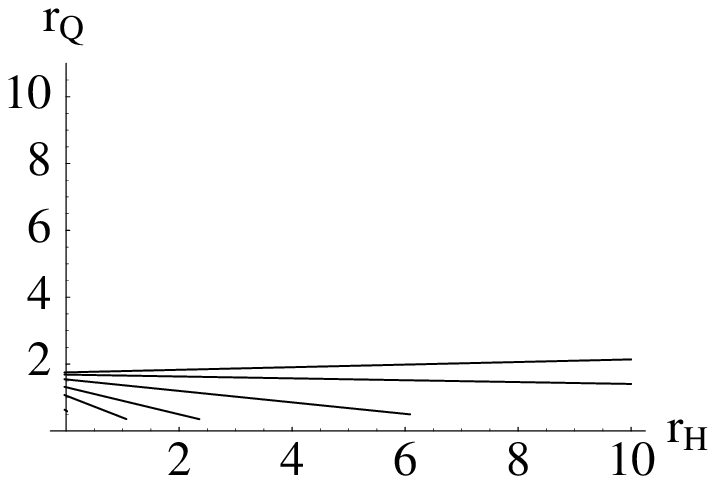}\hfill
\includegraphics[width=0.48\linewidth]{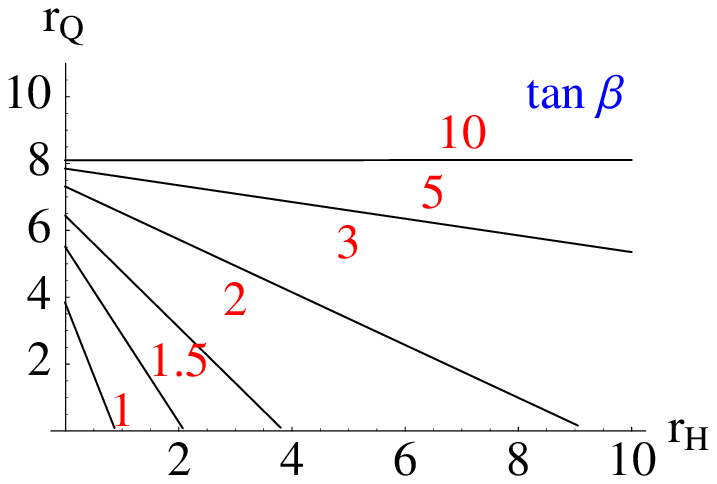}
\caption{The relation between $r_Q$ and $r_H$  necessary to
allow for tuning
of the weak scale, in the case of SU(5) GUT boundary conditions. We have
taken
$\mtil=10^6$ GeV ($10^{13}$ GeV) in the left (right) panels,
and $\tan\beta$ varying from 1 (lower line) to 10 (upper line).}
\label{region} 
\end{figure}

Let us consider the case in which the soft masses satisfy SU(5)
GUT boundary
conditions, ${\overline m}_{Q}={\overline m}_{U}={\overline m}_{E}$ and
${\overline m}_{D}={\overline m}_{L}$. This case can be realized when
supersymmetry is broken in a hidden sector and communicated at a scale
larger than $M_X$. Then, in terms of $r_Q={\overline m}_{Q_3}^2/
{\overline m}_{H_u}^2$ and $r_H={\overline m}_{H_d}^2/
{\overline m}_{H_u}^2$, eqs.~(\ref{tun1})--(\ref{tun4}) are equivalent 
to\footnote{When $t^2>(3-4K)(1-\omega)/[3\omega(1-4K)]$ and $K<1/4$,
the upper bound on $r_H$ disappears. This has no consequences for our
discussion.} 
\beq
2Kt^2r_Q=\left[ \omega \left( t^2+1\right) -1\right] r_H 
+t^2 \left( 1-K -\omega\right) -\omega 
\eeq
\beq
-\frac{\omega}{1-\omega}<r_H<\frac{t^2\left[ 3(1-\omega)-K(7-12\omega)
\right] -(3-4K)\omega}{(3-4K)(1-\omega)-3t^2\omega (1-4K)}.
\label{rangeh}
\eeq
The allowed region of $r_Q$--$r_H$ parameters is shown in \fig{region},
for characteristic values of $\mtil$ and $\tan\beta$. The tuning of
the electroweak scale can be achieved in a large area of parameters
for $r_Q$ and $r_H$ of order unity. Natural values of the boundary 
conditions are compatible with the breaking. 

As apparent from \fig{region}, the available area of $r_Q$--$r_H$
shrinks as $\tan\beta$ is lowered. Indeed, from \eq{rangeh}, we
obtain that a solution for $r_H$ exists only if
\beq
K<\frac{3}{7} \left( \frac{1-2\omega}{1-\frac{19}{7}\omega}\right) .
\label{upk}
\eeq
This bound is more stringent than the constraint $K<1/2$ (from
absence of Landau poles). Therefore \sps\ with SU(5) GUT
boundary conditions {\it cannot} have the top Yukawa coupling at the
infrared fixed point\footnote{The infrared fixed point of the top 
Yukawa coupling in \sps\ was recently studied in ref.~\cite{huitu}.}.
The origin of the bound in \eq{upk} can be easily traced.  
Larger values of $K$ require smaller
values of $r_Q$ to tune the weak scale, see \eq{tun1}. But if $r_Q$
is too small, \eq{tun4} can no longer be satisfied and the
stop mass square becomes negative.

\begin{figure}
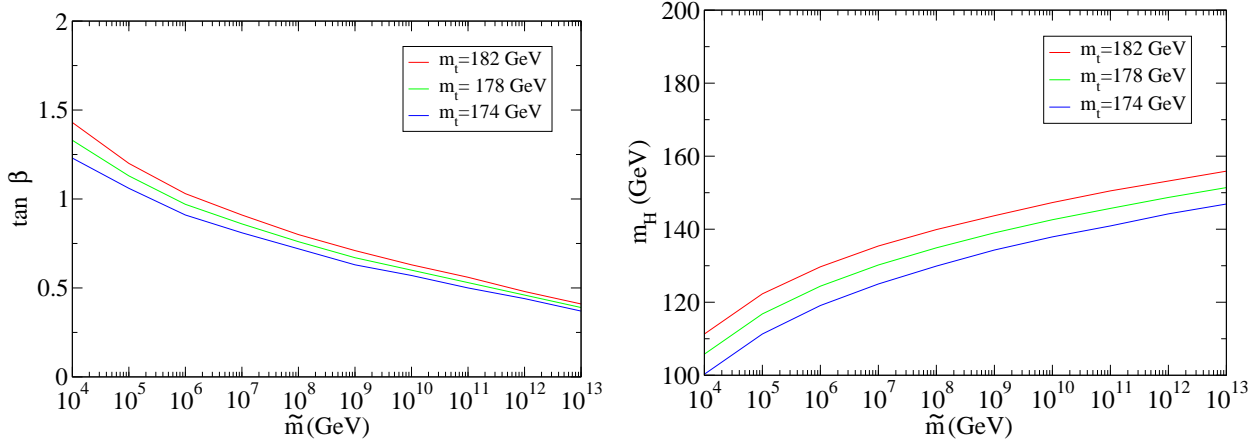

\centering
\includegraphics[width=0.48\linewidth]{tbeta.eps}\hfill
\includegraphics[width=0.5\linewidth]{higgs.eps}
\caption{Lower bound on $\tan\beta$ (left) and on $m_H$ (right)
obtained by requiring weak-scale tuning, in the case of SU(5) GUT
boundary conditions. The
different lines correspond to $m_t^{\rm pole}=174$, 178, 182 GeV.}
\label{upper} 
\end{figure}

An upper bound on $K$ corresponds to an upper bound on
$\lambda_t(\mtil)$, see \eq{cappa}. Since $\lambda_t(\mtil)=
h_t(\mtil)/\sin\beta$, \eq{upk} gives a lower bound on $\tan\beta$.
This bound in shown in \fig{upper}. We fix $\alpha_s(M_Z)=0.119$ and,
because of the significant
dependence on the top-Yukawa coupling, we show the result for
three values of the pole mass, $m_t=178\pm 4$  GeV. Since we are considering
general boundary conditions for the Higgs soft masses (\ie\ $r_H \ne
1$), $\tan\beta$ can in principle be smaller than 1. Low values of
$\tan\beta$ help in obtaining bottom-tau unification. The lower
limit on $\tan\beta$ can be translated into a lower limit on the Higgs
mass, as shown in \fig{upper}.

We also remark that SU(5) GUT boundary conditions allow for the possibility
of a double fine-tuning in which both the weak scale and the charged
Higgs mass ($m_{H_u}^2+m_{H_d}^2$) are much smaller than $\mtil$, and
therefore the low-energy theory has two Higgs doublets. This is obtained
for 
\beq
r_H=-\frac{\omega}{1-\omega},~~~~r_Q=\frac{1-K-\omega (2-K)}{2K(1-\omega)},
\eeq
which corresponds to the border line of the area in \fig{region} at low
$r_H$. However, the double fine-tuning requires a peculiar boundary
condition, with $r_H$ very small and negative.

\begin{figure}
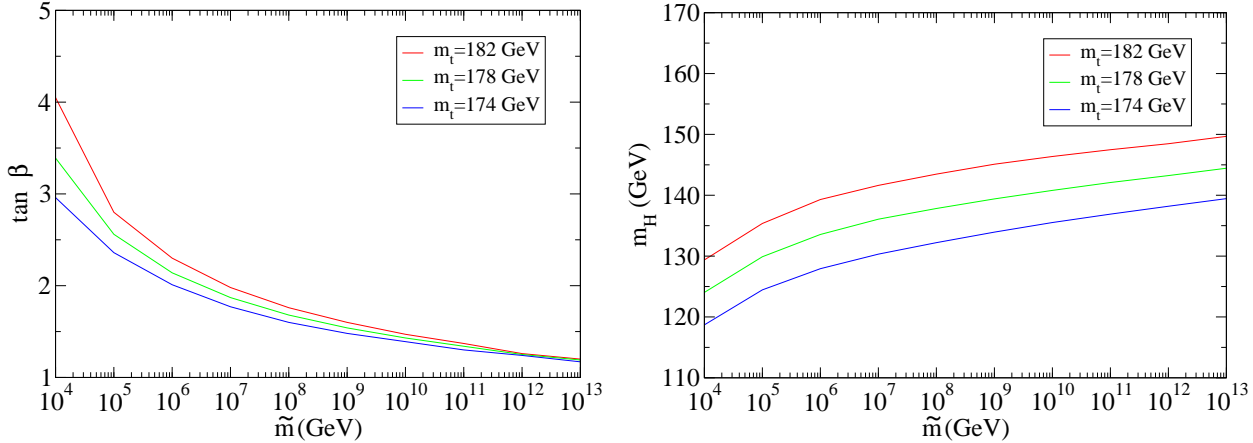

\centering
\includegraphics[width=0.48\linewidth]{tbetauni.eps}\hfill
\includegraphics[width=0.5\linewidth]{higgsuni.eps}
\caption{Values of $\tan\beta$ (left) and $m_H$ (right)
obtained by requiring weak-scale tuning, in the case of universal
boundary conditions. The
different lines correspond to $m_t^{\rm pole}=174$, 178, 182 GeV.}
\label{universal} 
\end{figure}

With stronger assumptions on the soft masses, we can obtain sharper
predictions. Let us consider the universality hypothesis, which implies
$r_Q=r_H=1$. In {\sps}, universality is not necessary to solve the flavour
problem, but it could appear in particular schemes of supersymmetry
breaking. In this case, the tuning of the weak scale leads to a prediction
of the value of $\tan\beta$
\beq
\tan\beta =\frac{1}{\sqrt{1-3K}},
\eeq
which is shown in \fig{universal}. In \sps\ with universal boundary
conditions, the Higgs mass is
uniquely predicted, as a function of $\mtil$, as shown in \fig{universal}. The
band in \fig{universal} corresponds to the uncertainty in $m_t$ and therefore the
prediction will be further sharpened, if the experimental error in the
top mass is reduced.

Finally, it is important to mention that, if we abandon SU(5) GUT boundary
conditions and consider the most general pattern of soft terms, the
lower bound in \eq{upk} can be evaded and the top-Yukawa infrared
fixed point can be reached. Indeed, for arbitrary soft terms,
even when ${\overline S}=0$, we
can tune the weak scale for any value of $\tan\beta$ and any value
of $K$ (with $0<K<1/2$) as long as $r_H<t^2(1-2K)/(1-K)$.

The reason why the tuning of the weak scale can be obtained so easily
for natural initial values of the soft masses and even for universal
boundary conditions lies on the efficiency of the radiative breaking
in supersymmetry and, ultimately, in the favourable observed value
of the top mass. The possibility of reproducing the weak scale for 
large scalar soft masses with universal boundary conditions
(first observed in ref.~\cite{barbieri})
was carefully studied in ref.~\cite{feng}, where it was named ``focus
point''. \sps\ brings that possibility to the extreme consequences.

In conclusion, we have discussed the meaning of the tuning condition 
in \sps\ and shown how it leads
to interesting information both on the underlying soft masses and
on low-energy measurable observables, like the Higgs mass $m_H$
and $\tan\beta$. 
Given specific high-energy matching conditions, one can make testable 
predictions. In particular, we have shown that SU(5) GUT boundary conditions
imply a lower bound on $\tan\beta$ and on $m_H$, while universal
boundary conditions imply a determination of $\tan\beta$ and $m_H$,
as functions of $\mtil$. Viceversa, measurements of the low-energy
parameters can help us discriminating among underlying models of
supersymmetry breaking.

We thank N.~Arkani-Hamed, R.~Rattazzi and A.~Romanino 
for very useful discussions.


\end{document}